# Dominant chiral optical forces in vicinity of optical nanofibers

M.H. ALIZADEH,[1,2,*] B.M. REINHARD,[1,2,*]

[1]Photonics Center, Boston University—8 Saint Mary Street, Boston, Massachusetts 02215, USA
[2]Department of Chemistry, Boston University, 590 Commonwealth Avenue, Boston, Massachusetts 02215, USA
*Corresponding author: halizade@bu.edu , bmr@bu.edu

**Transverse spin angular momentum (SAM) of light and associated transverse chiral optical forces have received tremendous attention recently as the latter may lead to an optical separation of chiral biomolecules. In this context, the relative magnitude of chiral and non-chiral forces represents a challenge for the implementation of chiral separation schemes. In this work we demonstrate that, by spatially separating the maxima of transverse spin density from the gradient of field intensity, it is possible to dominate chiral-specific components of the force over non-chiral ones. To that end, we study optical nanofibers and nanowires as a candidate for such scheme and demonstrate that in their vicinity, chiral optical forces can emerge that are stronger than gradient and scattering forces. This finding may be of significance in the design of improved optical separation schemes for chiral biomolecules.**

http://dx.doi.org/

Transverse spin angular momentum (SAM) of evanescent and structured electromagnetic (EM) fields has recently attracted considerable attention [1-9]. Unlike the SAM of propagating fields, this novel SAM is transverse to the direction of the wavevector. SAM of light can be written as [1, 10]:

$$S=S_e+S_m \quad, \quad S_e=\frac{\varepsilon_0}{4\omega}\Im m(E\times E^*), \quad S_m=\frac{\mu_0}{4\omega}\Im m(H\times H^*).$$ (1)

It is easily verified that transverse SAM is non-zero only when the field is elliptical, and for propagating waves is either parallel or anti-parallel to the direction of the propagation. However, when there is a nonzero longitudinal field component with a phase shift with respect to the transverse component of the field, the transversality condition: $k.E = 0$, dictates the existence of a spin component in the transverse direction to the propagation. This set of conditions can be met in different types of EM fields associated with, for instance, evanescent waves in planar interfaces [11], surface plasmon polaritons, non-paraxial beams [1, 12, 13] and as was recently shown in optical nanofibers and semiconductor nanowires [9]. Lateral chiral optical forces that result from transverse SAM are one of the major motivations behind the recent surge in interest in transverse SAM. Such lateral chiral forces appear in opposite directions for chiral objects of opposite handedness (enantiomers) [2, 14-27]. In the absence of strong gradients of optical chirality density, the chiral-specific optical force is proportional to the density of SAM [14, 28]:

$$F_{chiral} \propto \Im m[\gamma^*(\alpha+\beta/c^2)]\langle S\rangle$$ (2)

where $\alpha$ is the electric polarizability, $\beta$ is the magnetic polarizability and $\gamma^*$ is the complex conjugate of chiral polarizability (see Supplement 1). A big challenge for practical applications of chiral optical forces is that, in most cases their magnitude is much smaller than the EM gradient forces, which makes their detection challenging. This is due to the fact that locations of non-trivial transverse spin density also exhibit large electric field intensity. Here, we demonstrate that by spatially separating the maxima of spin density from those of the gradient of field intensity, chiral optical forces can emerge that are stronger than both gradient and scattering forces. In particular, we show that the waveguide modes of nanowires and nanofibers that have dominantly magnetic characteristic, can lead to a distribution of transverse SAM that peaks in the spatial minima of the electric field intensity, where it generates lateral chiral optical forces that dominate over gradient or scattering forces. We first investigate the spin properties of the fundamental mode of a subwavelength silicon nanowire and then use full-wave EM simulations to calculate the optical forces acting on a pair of chiral enantiomers in the vicinity of these structures. Our results reveal that when the chiral enantiomers reside in the maxima of the transverse SAM, chiral forces can exceed non-chiral gradient and scattering forces. As the lateral size of a cylindrical nanowire gets smaller it reaches a limit in which only the fundamental waveguide mode, i.e. the $HE_{11}$ mode, can be sustained in the structure [29, 30]. This lateral size confinement, naturally leads to a delocalization of the mode intensity into the ambient space. This modal behavior has been extensively employed for optical trapping of atoms and ions in the vicinity of optical nanofibers [31]. For nanofibers and nanowires with a deeply subwavelength radius, the fundamental mode intensity has a sinusoidal azimuthal dependence with field discontinuities on the radial boundary [29, 32, 33]. This mode, despite being guided along the nanowire long axis, is weakly confined laterally and has a large evanescent tail in the nanowire surroundings. This modal behavior was tested by exciting the $HE_{11}$ mode in a silicon nanowire, with a radius of 50 nm and length of 3μm, under linearly polarized incident light. It is clearly seen in Fig. 1, where the electric and magnetic field distributions, as well as their vectorial behaviors are shown, that the electric field intensity maxima of the mode point in the direction of the polarization of the incident light. Importantly, the mode exhibits large mode delocalization into the surrounding ambient medium, Figs. 1(b),1(c). The evanescence of the lateral field combined with the Gauss's law for charge-free surfaces written in k space, indicates a π/2 phase shift between $E_ρ$ and $E_z$, which leads to a nontrivial ellipticity of $HE_{11}$ mode in ρz plane inducing a spin in the azimuthal direction, $\hat{\phi}$, which is normal to the **k** vector [10, 29]. This is seen in Fig. 1(d), where the electric field lines around the nanowire are plotted for λ= 600 nm, at which wavelength, the strongest coupling of the incoming light to the $HE_{11}$ mode occurs. The electric field shows a distinct ellipticity, similar to a spinning wheel, that leads to transverse SAM directed normal to both the plane of rotation and the direction of **k**. The emergence of transverse spin is directly connected with excitation of a waveguide mode with an evanescent tail. When the incoming light does not couple to such a mode, the ellipticity of the mode is trivial and no appreciable transverse spin is generated. Spatial maps of the spin reveal an azimuthally dominant vectorial character and peaks in the plane normal to the polarization axis. The lack of any radial or z component of the spin vector is due to the fact that the incident light is linearly polarized and lacks any spin along the Poynting vector. Also, the larger spin density in the xz plane originates from the magnetically dominant nature of the HE modes. The magnetic field intensity maxima lie in the xz plane, perpendicular to the direction of the incident polarization. Since SAM is a sum of electric and magnetic contributions, the resulting spin density has an azimuthal distribution with a larger magnitude in the xz plane. This is of great significance when chiral optical forces are concerned. One persistent caveat of previously proposed chiral optical force schemes, based for instance on evanescent waves at dielectric interfaces or surface plasmon polaritons [15, 17], has been the dominance of gradient and scattering forces over the chiral optical forces. This shortcoming arises mainly due to the spatial overlap of the maxima of field intensity and transverse spin density. This is in sharp contrast with the transverse spin emerged in nanowires. As we have demonstrated the largest spin densities occur in a plane normal to the maxima of the electric field intensity. This behavior facilitates the generation of large transverse spin densities at locations that do not contain significant field gradients. As a result, the chiral optical forces will have similar or even larger magnitude than that of the gradient and scattering forces. Some other proposals have been based on taking advantage of gradient of optical chirality to separate or trap chiral particles [18, 20, 23, 24]. They suffer from the same limitation that locations with large gradient of optical chirality inevitably incorporate significant gradients of field intensity, which leads to a dominance of non-chiral gradient forces over chiral forces. In the proposed approach, however, due to the fact that the system of incident light plus the nanowire is non-chiral, no net optical chirality is generated. Also, since the maxima of electric and magnetic fields lie in perpendicular planes, the local density of optical chirality is insignificant. We confirmed this hypothesis by calculating the electromagnetic forces exerted on a pair of chiral structures placed in the maxima of the transvers spin density near a silicon nanowire. To this end, we employed a silicon nanowire with a radius of 55 nm and a length of 3 μm. The radius and the length of the nanowire were chosen to make sure that only the fundamental mode is excited and the predicted modal behavior holds true for the designated nanowire. For the denoted dimensions, the fundamental mode is optimally excited around λ=600 nm. We then placed a pair of silver chiral helices in the vicinity of the nanowire. The size of the helices was chosen to mimic the ultra-subwavelength length scale of chiral biomolecules. (see Supplement 2). We employed Maxwell's Stress Tensor method to calculate the optical forces exerted on the chiral enantiomers in the spectral range from 500 – 900 nm (Fig. 2). Different components of the optical forces are plotted separately in Fig. 2. In gradient and scattering forces two conspicuous peaks are observed, one of which originates from the excitation of the fundamental waveguide mode, while the other stems from the plasmonic resonances of the helices. We first focus on the observed peaks in $F_x$. Fig. 2b contains the calculated $F_x$ for the right-handed and left-handed helices, denoted as $F_{xr}$ and $F_{xl}$ respectively. The first peak occurs at λ= 600 nm, the wavelength at which the optimal excitation of the $HE_{1,1}$ fundamental mode occurs. The gradient of the evanescent tail of the $HE_{1,1}$ mode points toward the nanowire, i.e. -x direction. Since the excitation wavelength of the fundamental mode is independent of the plasmonic resonance of the helices, the ensuing gradient forces occur at the same wavelength for both enantiomers, i.e. around λ=600 nm. The optical gradient forces at this wavelength, exhibit a narrow resonance bandwidth, which is the characteristic of low-loss photonic modes. At the plasmon resonance wavelength both enantiomers experience stronger gradient forces and the bandwidths of the gradient forces are broadened. The excitation of the plasmon resonance results in enhanced field intensities at both ends of the helix. Due to higher effective refractive index, this intensity is larger at the side pointing towards the silicon nanowire, which results in a net gradient force toward the nanowire. Due to the plasmonic origin of the field, the resulting optical forces are broad and the slight shift in the peaks of the plasmonic gradient forces for two enantiomers is due to the small spectral difference between the plasmonic resonances of the enantiomers. (see Supplemental 3) The $F_y$ plots in Fig. 2c reveal several intriguing features. Most importantly, for an achiral particle this component would cease to exist and it is non-zero only for chiral particles. As we had anticipated, the emergence of transverse SAM of the fundamental mode leads to chiral optical forces pointing in opposite directions for two enantiomers. Eq. (2) relates these chiral forces to spin density and since this density is maximally generated upon excitation of the fundamental mode, the peaks of the chiral forces appear at around λ = 600 nm, with no correlation with the plasmon resonances of the helices. For wavelengths larger than 600 nm there is an insignificantly small residual transverse force that results from the small evanescent tails of the waves traveling along the nanowire. At the $HE_{1,1}$ peak wavelength, the chiral optical forces have larger magnitudes than gradient and scattering forces. This is the main finding of this paper. The dominance of the chiral optical forces under these conditions is explained by two major factors. First, as was discussed earlier, the density of transverse SAM is largest in the plane normal to the plane of incident light. Second, the polarization of the incident light creates the largest field gradients in the direction of the polarization and, thus, in a plane orthogonal to the plane that contains the chiral force maxima. This spatial separation of the maxima of transverse spin density and the field gradient lies at the heart of our approach to achieve dominant chiral optical forces. The weakest component of all optical forces, is the scattering force, which results from the momentum transfer from the incident light. The largest momentum of the light occurs when the photonic mode of the nanowire is excited. The scattering force, however, is almost an order of magnitude smaller than the other two force components. Also, as expected, they appear in the same direction for both enantiomers. In summary, we have shown that the EM field delocalization of the fundamental mode of nanowires and optical fibers with subwavelength lateral dimensions, results in large densities of transverse SAM in the vicinity of these structures. This transverse spin adds a new degree of freedom to explore in the applications of nanowires and optical fibers. In particular, we demonstrated that transverse SAM in these structures can generate large chiral optical forces that point in opposite directions for two enantiomers and whose magnitude is larger than that of gradient and scattering forces. These findings provide a rational design approach for the next generation of photonic traps capable of an optical enantio-selective separation of chiral biomolecules.

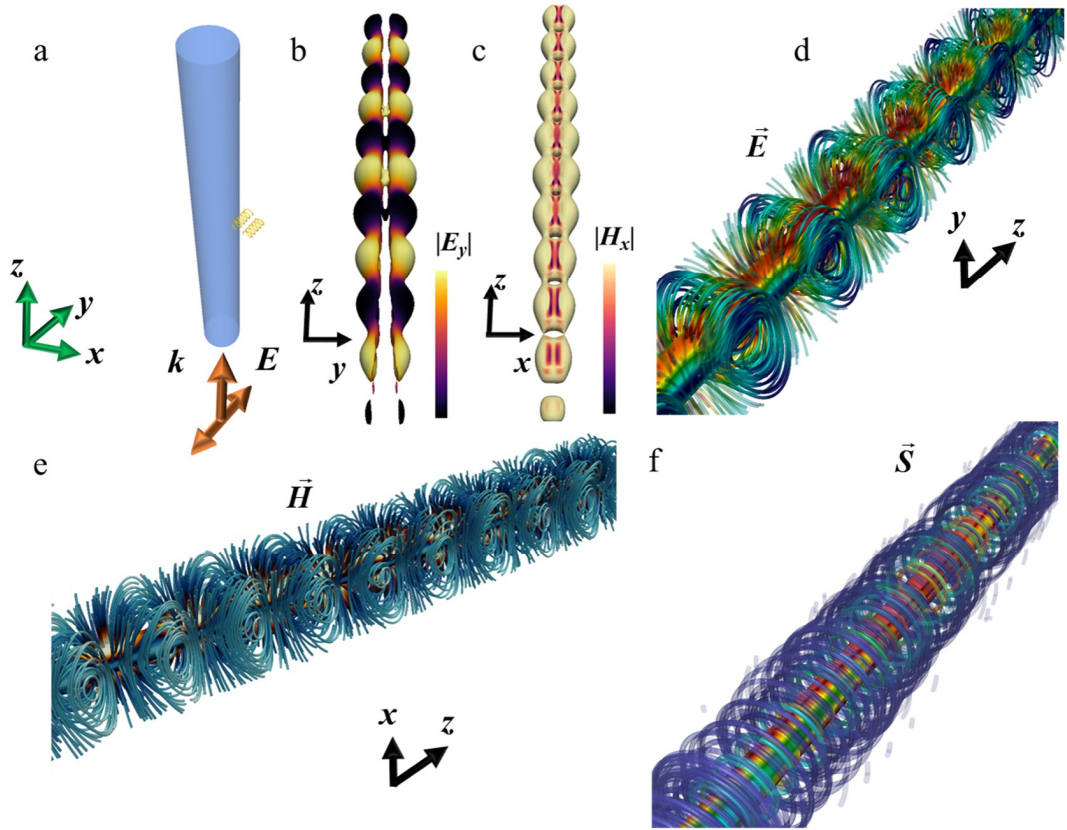

**Fig. 1.** (a) The schematics of excitation of the fundamental mode of the silicon nanowire (b) The electric field distribution in the vicinity of the nanowire is shown for λ=600 nm. The electric field peaks along the polarization axis of the incident light in y direction The most relevant feature of the HE11 mode, which is its significant modal spill-over to the surrounding, is clearly seen in this plot. (c) Magnetic field distribution for the $HE_{11}$ mode. The peaks occur perpendicular to the polarization axis, i.e. in x direction. (d) The distribution of the electric field lines around the nanowire. The mode travels along the nanowire with a large evanescent tail. The spinning behavior of the electric field of the $HE_{11}$ mode which gives rise to the transverse spin is clearly seen. (e) The distribution of the magnetic field lines around the nanowire for λ=600 nm. The spinning magnetic field provides the major contribution to the transverse spin. (h) A bird's eye view of the azimuthal distribution of SAM is shown around the nanowire surface for λ=600 nm.

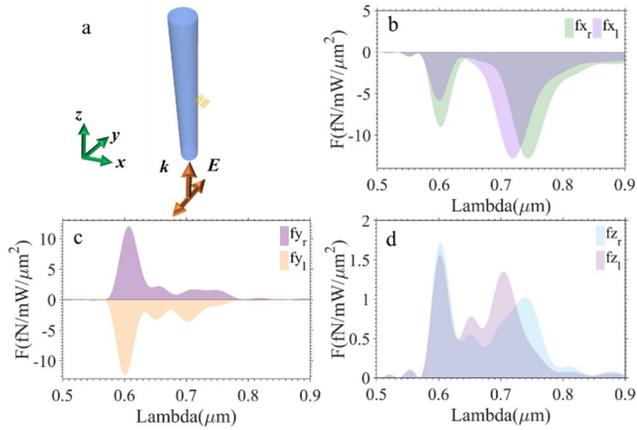

**Fig. 2.** Calculated electromagnetic forces for a pair of chiral enantiomers near a silicon nanowire. (a) Excitation geometry of the fundamental mode in the silicon nanowire. This mode generates a large density of transverse spin in the plane perpendicular to the plane that contains polarization vector and the wavevector, which leads to dominant transverse chiral forces. (b) The x components of the forces are plotted for right-handed and left-handed enantiomers. The x component represents the gradient force, which is directed towards the nanowire. (c) The transverse components of the forces acting on the two chiral enantiomers. The y component of the forces is expected to vanish for non-chiral objects, but the transverse spin results in significant transverse chiral forces that peak along the y direction. (d) The z component of the optical force results from the scattering force and is significantly smaller than the other two forces.

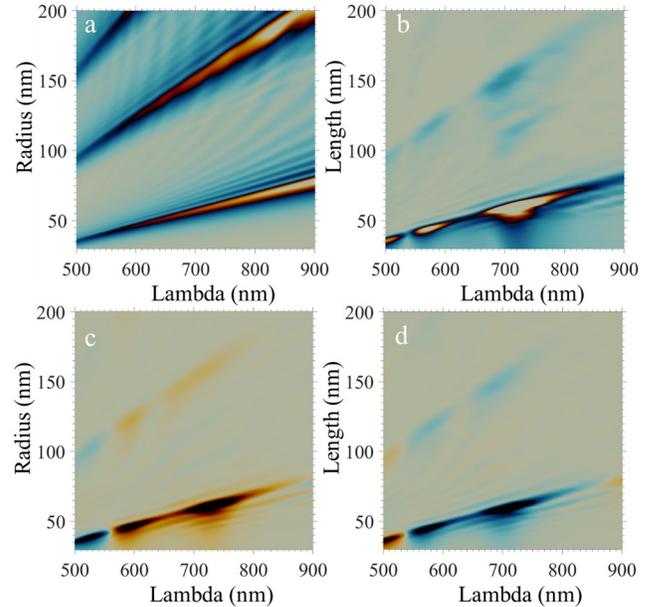

**Fig. 3.** Chiral optical forces in higher modes of the nanofiber. (a) Scattering cross section plot for silicon nanowires, with a length of 3 μm and increasing radius, from 30 nm to 200 nm. For radii larger than 100 nm, higher optical modes can be sustained in the structure. (b) calculated total force for the left-handed helix in the vicinity of the nanowire. Since all the force components directly depend on excitation of a strong evanescent field in the vicinity of the nanowire, the total force is significant only for the fundamental mode. (c), (d) More interestingly, the transverse chiral forces are non-trivial merely for the fundamental mode and almost vanish for higher modes. These results confirm that the modal spill-over of the fundamental mode has an underlying role in emergence of chiral optical forces in optical nanofibers and nanowires.

**Funding:** This work was supported by a Boston University Materials Science and Engineering Innovation Award.

# Appendix

## 1. Derivation of equation 1 in the main text.

The electromagnetic force on a dipolar particle can be written as [14, 34]:

$$\langle F \rangle = \frac{1}{2}\Re[(\nabla E^*).p + (\nabla H^*).m - \frac{ck_0^4}{6\pi}(p \times m^*)] \quad (S1)$$

where **p** and **m** are the electric and magnetic dipoles, respectively. A chiral dipole is distinct from an achiral counterpart in that, a cross-polarization term connects the electric dipole moment to the magnetic field and the magnetic dipole moment to the electric field. More specifically the electric and magnetic dipole moments for a chiral dipole are defined as:

$$\begin{bmatrix} p \\ m \end{bmatrix} = \begin{bmatrix} \alpha & i\gamma \\ -i\gamma & \beta \end{bmatrix} \begin{bmatrix} E \\ H \end{bmatrix} \quad (S2)$$

where $\alpha$ is the complex electric polarizability, $\beta$ is the complex magnetic polarizability and $\gamma$ is the complex chiral polarizability. By substituting Eq. (S2) in Eq. (S1) and assuming that $|\Re\gamma| << |\Im\gamma|$ one can obtain the force on a chiral dipole as [28]:

$$\langle F \rangle = F_{chiral} + F_{Achiral}$$

$$F_{chiral} = -\Re(\gamma)\nabla \frac{1}{2}\Im[E^*.B] - \frac{k^5}{3\pi\varepsilon_0^2}\Im\left[\gamma^*(\alpha + \frac{\beta}{c^2})\right]\langle S \rangle$$

$$F_{Achiral} = \left[\nabla \frac{1}{4}\Re(\alpha)|E|^2 + \frac{1}{4}\Re(\beta)|H|^2\right] + \left(\frac{C_{ext} + C_{recoil}}{c}\right)\langle \Pi \rangle + C_{ext}c\nabla \times \langle S \rangle$$

(S3)

where $k_0$ is the wavenumber of the incident light, $\langle S \rangle$ is the spin density, $C_{ext}$ and $C_{recoil}$ are the extinction and recoil cross sections, respectively, and $\langle \Pi \rangle$ is the time-averaged Poynting vector. A closer look at the electromagnetic force in Eq. (S3) reveals that it comprises a chiral and an achiral part. The achiral force has three major terms. The first term, is the gradient force, which is normally dominated by the gradient of the electric field intensity, as $\Re(\alpha) >> \Re(\beta)$. The second term is the scattering force, which is proportional to the Poynting vector and the last term is spin-momentum force, which is proportional to the curl of spin, which is sometimes referred to as Belinfante momentum. The mentioned forces are not chiral specific, as they do not depend on chiral polarizability. The chiral part of the optical force, however, depends
on chiral polarizability, $\gamma$, and since it changes sign for chiral particles of opposite handedness, the resulting forces will change sign as well. The first term of the chiral force, is directly proportional to the gradient of the optical chirality. For the density of optical chirality to be non-zero, complex electric and magnetic fields should have phases-shifted parallel component. For instance, this condition can be met for circularly polarized light. Even when the density of optical chirality is non-zero, the chiral gradient-force gets dominated by the achiral gradient force. This is mainly because the loci with non-trivial optical chirality are the ones with high density of electric field. In our case, however, net optical chirality produced in the vicinity of the optical nanofiber is zero. This stems from the fact that both the incident light and the photonic structure are achiral. This results in the chiral force being dominated by the spin component. On the other hand, since $\beta << \alpha$, the second term in the chiral force due to spin, can be neglected which finally leads to:

$$F_{chiral} = -\frac{k^5}{3\pi\varepsilon_0^2}\Im(\gamma^*\alpha)\langle S \rangle \quad (S4)$$

The remaining question is to justify the role of transverse spin in the context of the chiral forces. In the lack of transverse spin, the chiral forces will be in the direction of the longitudinal spin. For the propagating electromagnetic waves, this is either parallel or antiparallel to Poynting vector. This makes the chiral forces, overshadowed by the scattering forces. In the presence of transverse spin, however, the resulting chiral forces will be in normal direction to both the gradient force and the scattering force. In our case, in particular, not only the direction of the generated chiral forces is perpendicular to the scattering and gradient forces, its magnitude is dominant over these forces.

## 2. Chiral polarizability of the simulated silver helix

The assumption that $|\Re\gamma| << |\Im\gamma|$ is justified for a plasmonic chiral particle and for any chiral bio-molecule at wavelength range close to its absorption band. In the case of a metal helical particle, for instance, one can find the chiral polarizability to be [15, 35]:

$$|\gamma| \approx \frac{l^2}{4\pi k}\left(1 - \frac{16a}{3\pi l} + \frac{2a^2}{l^2} - \frac{4a^4}{l^4}\right)^{-1}$$ where $l$ is the length of the helix,

$a$ is the radius and $k$ is the wavenumber. Also the electric polarizability can be estimated by: $\alpha \approx \frac{l^3}{4\pi^2 k^2 a^2}\left(1 - \frac{16a}{3\pi l} + \frac{2a^2}{l^2} - \frac{4a^4}{l^4}\right)^{-1}$. For the silver helix in our simulations $l$=36 nm, $a$=7.5 nm, and $\lambda$=600 nm, which yield $|\gamma| \approx 3.32 \times 10^3\, nm^3$ and $\alpha \approx 1.61 \times 10^6\, nm^3$.

## 3. Mie calculations for the helical enantiomers

The slight spectral shifts observed in the calculated optical force components stem from small difference in the Mie spectra of the helices. This is seen in Fig. (S1), where the Mie spectra for these structures are plotted. It is clearly observed that the resonance peaks are shifted the same way the peaks of the force components are in the main text.

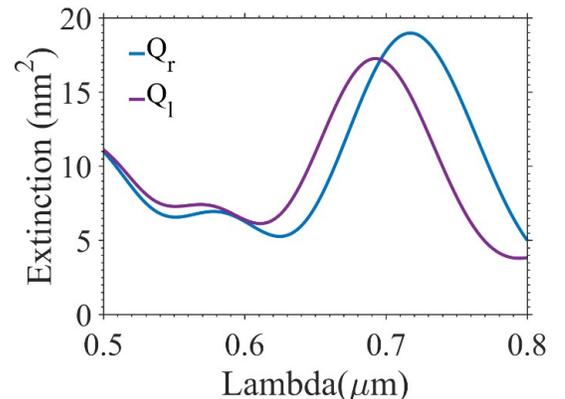

**Fig. (S1)** Mie spectra of the right handed and left handed silver helices.

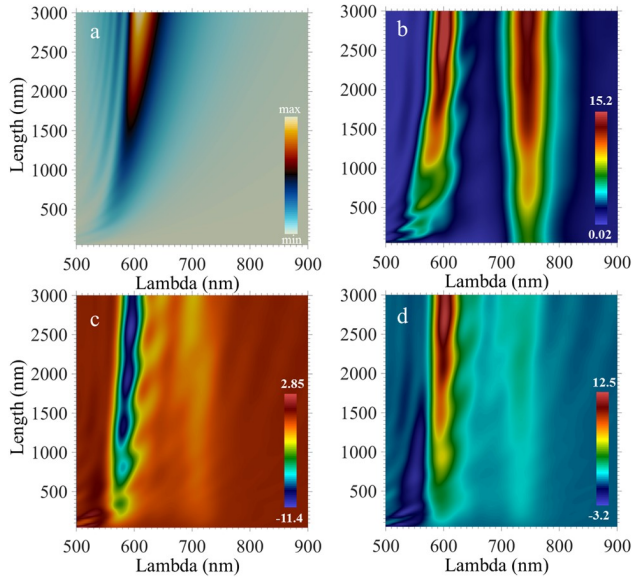

Fig. (S2). HE$_{11}$ mode dependence of chiral forces. (a) Scattering spectrum of the silicon nanowire with a radius of 55 nm as function of length between 50 nm to 3 µm. Due to the lateral size of the nanowire, only the fundamental mode can be excited. (b) Contour plot of the magnitude of the total optical force for the left handed helix as a function of the wavelength and length. The total force plot for the right handed helix is similar. (c) The chiral force for left handed helix. This force emerges only after the fundamental mode is efficiently excited and disappears when this mode is not excited, even at the plasmon resonance wavelength. (d) Same as in (c), but for the right-handed helix. It is clearly observed that the chiral force changes sign for the right-handed enantiomer. In all plots forces per area and excitation power are given in units of fN/mW/µm$^2$.

## 4. Dependence of the chiral optical force on HE11 mode

In order to verify our argument that the observed dominant chiral forces are due to the emergence of transverse SAM, we further analyzed the dependence of the chiral forces on the length of the nanowire. The assertion that the origin of the observed chiral optical forces is the transverse SAM of the optical mode of the nanowire, requires direct dependence of these forces on excitation of such modes. In other words, structures that are incapable of sustaining these waveguide modes will not generate chiral forces in their vicinity. An immediate example are silicon disks with the same radius as the investigated nanowire, but shorter length. These structures only sustain quasi-static Mie resonances that do not propagate and as a result do not generate any spin features in their immediate medium. As further verification, we systematically performed electromagnetic force calculations for chiral nanohelices in the vicinity of structures with subwavelength radius of 55 nm but with varying length [Fig. (S2)]. Fig. (S2.a) shows the scattering cross section for silicon structures of the same length range, i.e. radius of 55 nm and changing length between 50 nm and 3 µm. We chose such a vast wavelength range to span the relevant electromagnetic modal regimes, from Mie resonances to waveguide modes. As it is clear from Fig. (S2.a), the spectrum is dominated by the fundamental HE$_{11}$ mode which is excited around λ = 600 nm. This mode is excited only for lengths longer than a defined threshold, after which a propagating waveguide mode can be sustained in the structure and the momentum-matching condition can be met. The chosen lateral dimension of the structure does not allow for higher modes to be excited in the structure. The plots for chiral optical forces show an excellent agreement with the scattering spectrum. The chiral forces appear only for the silicon structure that are long enough to support the HE$_{11}$ mode and they emerge exactly around λ=600 nm. Since the chiral forces depend on the spin density, they vanish for any other wavelength, including those with large gradient force components. On the other hand, the total force, Fig. (S2b), experiences a second peak, which is due to the plasmon resonance of the helices.

## 6. Near-field map of the higher waveguide modes

Fig. (S4) demonstrates the near-field intensity distribution for the fundamental and the higher waveguide modes.

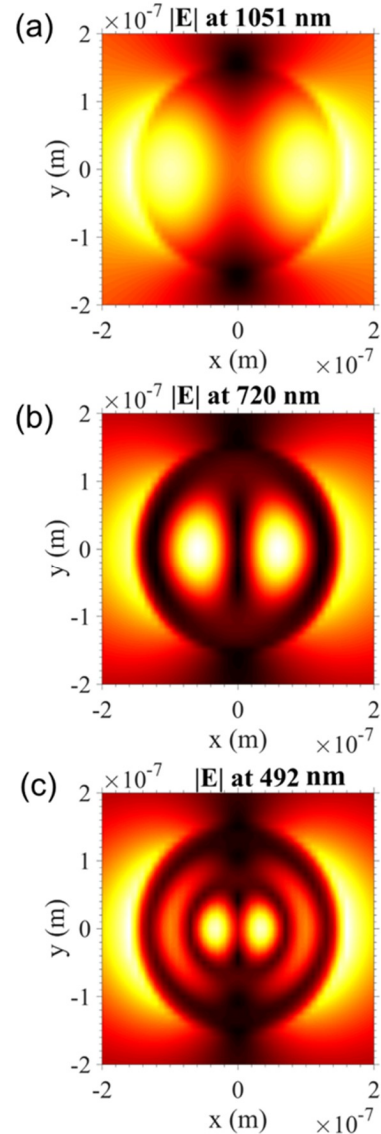

**Fig. (S4)** Field intensity distribution in the cross section of a nanowire with a radius of 150 nm and a height of 3 µm. (a) The HE$_{11}$ (b) HE$_{21}$ and (c) HE$_{31}$ modes.